\documentclass[11pt]{article}

\usepackage{palatino}
\usepackage{mathpazo}
\usepackage{stmaryrd}

\usepackage{braket}
\usepackage{amsfonts}
\usepackage{amssymb}
\usepackage{amsmath}
\usepackage{latexsym}
\usepackage{amsthm}
\usepackage{eepic}
\usepackage{sectsty}
\usepackage[usenames]{color}
\usepackage{hyperref}

\hypersetup{pdfpagemode=None}

\newtheorem{theorem}{Theorem}

\newtheorem{cor}[theorem]{Corollary}
\theoremstyle{definition}

\newcommand{\tinyspace}{\mspace{1mu}}

\newcommand{\op}[1]{\operatorname{#1}}

\newcommand{\norm}[1]{\left\lVert\tinyspace #1 \tinyspace\right\rVert}

\newcommand{\tr}{\operatorname{Tr}}

\newcommand{\ip}[2]{\left\langle #1 , #2\right\rangle}

\def\I{\mathbb{1}}

\newcommand{\setft}[1]{\mathrm{#1}}
\newcommand{\lin}[1]{\setft{L}\left(#1\right)}
\newcommand{\density}[1]{\setft{D}\left(#1\right)}
\newcommand{\unitary}[1]{\setft{U}\left(#1\right)}

\newenvironment{mylist}[1]{\begin{list}{}{
	\setlength{\leftmargin}{#1}
	\setlength{\rightmargin}{0mm}
	\setlength{\labelsep}{2mm}
	\setlength{\labelwidth}{8mm}
	\setlength{\itemsep}{0mm}}}
	{\end{list}}

\newenvironment{namedtheorem}[1]
	       {\begin{trivlist}\item {\bf #1.}\em}{\end{trivlist}}

\def\X{\mathcal{X}}
\def\Y{\mathcal{Y}}

\begin{document}

\title{\bf\LARGE
  Mixing doubly stochastic quantum channels with\\ 
  the completely depolarizing channel} 

\author{John Watrous\\[1mm]
  {\it\small Institute for Quantum Computing and School of Computer
    Science}\\ 
  {\it \small University of Waterloo, Waterloo, Ontario, Canada.}
}

\date{July 16, 2008}

\maketitle
  
\begin{abstract}
  It is proved that every doubly stochastic quantum channel that is 
  properly averaged with the completely depolarizing channel can be
  written as a convex combination of unitary channels.
  As a consequence, we find that the collection of channels
  expressible as convex combinations of unitary channels has non-zero
  Borel measure within the space of doubly stochastic channels.
\end{abstract}

\maketitle
  
\section{Introduction} \label{sec:introduction}

Let $\X$ and $\Y$ be finite-dimensional complex Hilbert spaces, and
let $\lin{\X}$ and $\lin{\Y}$ denote the sets of linear operators
acting on these spaces.
By a {\it super-operator} one means a linear mapping of the form
\begin{equation} \label{eq:super-operator}
  \Phi: \lin{\X} \rightarrow \lin{\Y}.
\end{equation}
Such a super-operator is said to be {\it admissible} if and only if
it is both completely positive and preserves trace.
In the usual (Schr\"odinger) picture of quantum information, the
admissible super-operators represent valid physical operations,
transforming a system having associated space $\X$ to one with
associated space $\Y$.
For this reason, admissible super-operators are also commonly referred
to as {\it channels}.

A super-operator $\Phi$ is said to be {\it unital} if it is the case that
$\Phi(\I_{\X}) = \I_{\Y}$, for $\I_{\X}$ and $\I_{\Y}$ denoting the
identity operators on $\X$ and $\Y$, respectively.
Super-operators that are both unital and admissible are said to be 
{\it doubly stochastic}.
The existence of a doubly-stochastic channel of the form
\eqref{eq:super-operator} obviously requires that $\X$ and $\Y$ have
equal dimension $d$, which is assumed hereafter in this paper.

Among the doubly stochastic channels are the
{\it mixed-unitary}\footnote{%
  The term {\it random unitary} channel is commonly used.
  However, this term is easily confused with a different notion that
  is now common in quantum information theory---that of a {\it fixed}
  unitary operator that is randomly chosen according to some measure
  (frequently the Haar measure).
  For this reason, the term {\it mixed-unitary} channel is suggested
  in this paper.}
channels:
those for which there exists a collection of unitary operators
$U_1,\ldots,U_N\in\unitary{\X,\Y}$ and a probability distribution
$(p_1,\ldots,p_N)$ so that 
\begin{equation} \label{eq:random-unitary}
\Phi(X) = \sum_{i = 1}^N p_i U_i X U_i^{\ast}
\end{equation}
for all $X\in\lin{\X}$.
(The notation $\unitary{\X,\Y}$ refers to the set of unitary mappings
from $\X$ to $\Y$.)

For the case $d=2$ it holds that every doubly stochastic
channel is mixed-unitary, but for larger $d$ this is no longer the
case: for all $d\geq 3$ there exist doubly stochastic channels that
are not mixed-unitary \cite{Tregub86,KuemmererM87,LandauS93}.
While it is computationally simple to check that a given
channel is doubly stochastic, there is no efficient procedure
known to check whether a given channel is mixed-unitary.

Several papers have studied properties of mixed-unitary channels
within the last several years.
For instance, Gregoratti and Werner \cite{GregorattiW03} gave a
characterization, with an interesting physical interpretation,
of the class of mixed-unitary channels: they proved that a given
channel is mixed-unitary if and only if it has full 
{\it quantum corrected capacity}, which assumes that the correction
procedure is permitted arbitrary measurements on the channel's
environment.
Smolin, Verstraete, and Winter \cite{SmolinVW05} further investigated
this notion of capacity, and conjectured that many copies of any
doubly stochastic channel can be closely approximated by a
mixed-unitary channel.
Buscemi \cite{Buscemi06} investigated bounds on the number $N$
required in the expression \eqref{eq:random-unitary} for different
channels; Audenaert and Scheel \cite{AudenaertS08} investigated
conditions under which channels are mixed-unitary; and 
Mendl and Wolf~\cite{MendlW08} proved several facts about
mixed-unitary channels, including the fact that a non-mixed-unitary
channel can become mixed-unitary when tensored with the completely
depolarizing channel.
Rosgen \cite{Rosgen08b} recently proved that various additivity
questions about general channels reduce to the same questions on the
(potentially simpler) class of mixed-unitary channels.

This paper proves a simple fact about mixed-unitary channels, which is
that every doubly stochastic channel becomes mixed-unitary when
properly averaged with the completely depolarizing channel.
To state this fact more precisely, let us write
$\Omega:\lin{\X}\rightarrow\lin{\Y}$ to denote the completely
depolarizing channel, which is defined as
\[
\Omega(X) = \frac{\tr(X)}{d}\I_{\Y}
\]
for every operator $X\in\lin{\X}$.
This channel is well-known to be mixed-unitary, as it is representable
as a uniform mixture over the discrete Weyl operators (also known as
the generalized Pauli operators).
The claimed fact is now stated in the following theorem.

\begin{theorem} \label{theorem:main1}
  Let $\X$ and $\Y$ be $d$ dimensional complex Hilbert spaces,
  and let $\Phi:\lin{\X}\rightarrow\lin{\Y}$ be any doubly stochastic
  channel.
  Then for $0\leq p\leq 1/(d^2 - 1)$ it holds that
  \[
  p \, \Phi + (1 - p) \, \Omega
  \]
  is a mixed-unitary channel.
\end{theorem}

It is also proved, as a corollary of this theorem, that within the
smallest real affine subspace of super-operators that contains the doubly
stochastic channels, there is a ball with positive radius around the
completely depolarizing channel within which all super-operators are
mixed-unitary channels.

\section{Notation and background information}

The purpose of this section is to introduce concepts and notation that
are needed in the proof of the main result that appears in the
following section.

Assume throughout the remainder of the paper that $\X$ and $\Y$ are
finite-dimensional complex Hilbert spaces for which a common
orthonormal basis $\{\ket{1},\ldots,\ket{d}\}$ has been fixed.
We let $\lin{\X,\Y}$ denote the set of all linear operators from $\X$
to $\Y$, and (as stated above) the notation $\lin{\X}$ is shorthand
for $\lin{\X,\X}$ (and likewise for other spaces in place of $\X$).
The usual (Hilbert--Schmidt) inner product of two operators $X$ and
$Y$ is defined as
\[
\ip{X}{Y} = \tr(X^{\ast} Y).
\]

With respect to the standard basis of $\X$ and $\Y$, we define a
linear bijection
\[
\op{vec}:\lin{\X,\Y}\rightarrow \Y\otimes\X
\]
by setting
\[
\op{vec}(\ket{i}\bra{j}) = \ket{i} \otimes \ket{j}
\]
for $1\leq i,j\leq d$, and extending to all of $\lin{\X,\Y}$ by
linearity.

The {\it Choi-Jamio{\l}kowski representation}
\cite{Jamiolkowski72,Choi75} of a super-operator
$\Phi:\lin{\X}\rightarrow\lin{\Y}$ is defined as
\[
J(\Phi) = \sum_{1\leq i,j\leq d} \Phi(\ket{i}\bra{j}) \otimes
\ket{i}\bra{j}.
\]
The mapping $J$ defined in this way is a linear bijection from the
space of all super-operators of the form $\lin{\X}\rightarrow\lin{\Y}$
to the space $\lin{\Y\otimes\X}$.
It holds that a super-operator $\Phi:\lin{\X}\rightarrow\lin{\Y}$ is
completely positive if and only if $J(\Phi)$ is positive semidefinite,
trace-preserving if and only if $\tr_{\Y}(J(\Phi)) = \I_{\X}$, and
unital if and only if $\tr_{\X}(J(\Phi)) = \I_{\Y}$.
The following facts may be verified directly:
\begin{mylist}{\parindent}
\item[1.]
  For any choice of an operator $A \in \lin{\X,\Y}$, the
  super-operator $\Phi:\lin{\X}\rightarrow\lin{\Y}$ defined by 
  $\Phi(X) = A X A^{\ast}$ satisfies 
  $J(\Phi) = \op{vec}(A) \op{vec}(A)^{\ast}$.
\item[2.] 
  For any choice of operators $A\in\lin{\X}$ and
  $B\in\lin{\Y}$, the super-operator
  $\Phi:\lin{\X}\rightarrow\lin{\Y}$ defined by 
  $\Phi(X) = \ip{A}{X} B$ satisfies $J(\Phi) = B \otimes
  \overline{A}$.
\end{mylist}
When combined with the linearity of the mapping $J$, these facts will
allow for simple calculations of $J(\Phi)$ for particular
super-operators $\Phi$ appearing later in the paper.

Next, let us recall that the {\it swap operator} on $\Y\otimes\X$ is
defined as
\[
W = \sum_{1\leq i,j\leq d} \ket{i}\bra{j} \otimes \ket{j}\bra{i},
\]
and let us write $R$ and $S$ to denote the orthogonal projections on
$\Y\otimes\X$ defined as
\[
R = \frac{1}{2}(\I - W) \quad\text{and}\quad
S = \frac{1}{2}(\I + W).
\]
The projections $R$ and $S$ denote the {\it anti-symmetric} and
{\it symmetric projections} on $\Y\otimes\X$, respectively.
Subscripts are used when it is necessary to be explicit about the
spaces on which the projections $R$ and $S$ act; so, as just defined,
the operators $R$ and $S$ are written more precisely as
$R_{\Y\otimes\X}$ and $S_{\Y\otimes\X}$, and in general other spaces
may be substituted for $\X$ and $\Y$.
In the next section we will make use of the identities
\begin{equation} \label{eq:R-and-S}
\begin{split}
\tr_{\X} \left[(\I_{\Y} \otimes X) R \right] & = \frac{1}{2}\tr(X)
\I_{\Y} - \frac{1}{2} X,\\[2mm]
\tr_{\X} \left[(\I_{\Y} \otimes X) S \right] & = \frac{1}{2}\tr(X)
\I_{\Y} + \frac{1}{2} X,
\end{split}
\end{equation}
which hold for all $X\in\lin{\X}$.

Finally, it is necessary that a few points on integrals over unitary
operators are discussed.
Let us first note that the collection of mixed-unitary channels
$\Phi:\lin{\X}\rightarrow\lin{\Y}$ is convex and compact.
For any measure $\nu$ on the Borel subsets of
$\unitary{\X,\Y}$, normalized so that $\nu(\unitary{\X,\Y}) = 1$,
it follows that the super-operator defined as
\begin{equation} \label{eq:integral-channel}
\Phi(X) = \int U X U^{\ast}\, \mathrm{d}\nu(U)
\end{equation}
for every $X\in\lin{\X}$ is a mixed-unitary channel---or in other
words can be expressed in the form \eqref{eq:random-unitary} for some
finite choice of $N$.
(It can easily be shown, using Carath\'eodory's Theorem, that such an
expression exists for some choice of $N\leq d^4-2d^2+2$.
See also Buscemi \cite{Buscemi06} for bounds based on
$\op{rank}(J(\Phi))$.)

Hereafter let us write $\mu$ to denote the normalized Haar measure on
$\unitary{\X,\Y}$. 
This is the unique measure on the Borel subsets of $\unitary{\X,\Y}$
that satisfies $\mu(\unitary{\X,\Y})=1$ and is invariant under left
multiplication by every unitary operator $V\in\unitary{\Y}$.
It is clear that identity
\begin{equation} \label{eq:identity1}
\int \op{vec}(U) \op{vec}(U)^{\ast} \mathrm{d}\mu(U) 
= \frac{1}{d}\I_{\Y\otimes\X}
\end{equation}
holds.
We will also require the identity
\begin{align}
\int 
\left(
\op{vec}(U) \op{vec}(U)^{\ast} \otimes
\op{vec}(U) \op{vec}(U)^{\ast}
\right)
 \mathrm{d}\mu(U)\hspace{-4cm} \nonumber\\
& =\frac{2}{d(d-1)} R_{\Y_1\otimes\Y_2}\otimes R_{\X_1\otimes\X_2}
+ \frac{2}{d(d+1)} S_{\Y_1\otimes\Y_2} \otimes S_{\X_1\otimes\X_2},
\label{eq:identity2}
\end{align}
where $\X_1$ and $\X_2$ represent isomorphic copies of the space $\X$,
$\Y_1$ and $\Y_2$ are isomorphic copies of $\Y$, and we view that
\[
\int 
\left(
\op{vec}(U) \op{vec}(U)^{\ast} \otimes
\op{vec}(U) \op{vec}(U)^{\ast}
\right) \mathrm{d}\mu(U) \in
\lin{\Y_1\otimes\X_1\otimes\Y_2\otimes\X_2}.
\]
The identity \eqref{eq:identity2} may be confirmed by considering
the well-known {\it twirling} operation \cite{Werner89}
\[
\int \left(U \otimes U\right) X
\left(U \otimes U\right)^{\ast} d\mu(U)
= \frac{2}{d(d-1)} \ip{R}{X} R
+ \frac{2}{d(d+1)} \ip{S}{X} S,
\]
and taking the Choi-Jamio{\l}kowski representation of both sides.
(For the interested reader, it is noted that the above integrals
\eqref{eq:identity1} and \eqref{eq:identity2}, as well as higher-order
variants of them, may also be easily evaluated by means of a general
formula of Collins and \'Sniady \cite{CollinsS06}.)

\section{Proof of the main theorem and a corollary}

We now prove the main theorem, which is restated here for convenience.

\begin{namedtheorem}{Theorem~\ref{theorem:main1}}
  Let $\X$ and $\Y$ be $d$ dimensional complex Hilbert spaces,
  and let $\Phi:\lin{\X}\rightarrow\lin{\Y}$ be any doubly stochastic
  channel.
  Then for $0\leq p\leq 1/(d^2 - 1)$ it holds that
  \[
  p \, \Phi + (1 - p) \, \Omega
  \]
  is a mixed-unitary channel.
\end{namedtheorem}

\begin{proof}
Given that $\Omega$ is mixed-unitary, and that the set of
mixed-unitary channels is convex, it suffices to consider the
case $p = 1/(d^2 - 1)$.

Define a super-operator $\Psi:\lin{\X}\rightarrow\lin{\Y}$ as
\[
\Psi(X) = \int U X U^{\ast}
\ip{\op{vec}(U)\op{vec}(U)^{\ast}}{J(\Phi)} \mathrm{d}\mu(U).
\]
By the above identity \eqref{eq:identity1} we have that
\[
\int \ip{\op{vec}(U)\op{vec}(U)^{\ast}}{J(\Phi)} \mathrm{d}\mu(U) = 1,
\]
and it is clear that $\ip{\op{vec}(U)\op{vec}(U)^{\ast}}{J(\Phi)}$ is
nonnegative for all $U\in\unitary{\X,\Y}$.
It follows that $\Psi$ is a mixed-unitary channel.
To complete the proof, it will suffice to establish that
\begin{equation} \label{eq:Psi}
\Psi = \frac{d^2 - 2}{d^2 - 1} \Omega + \frac{1}{d^2-1} \Phi,
\end{equation}
for then the right-hand-side will be shown to be mixed-unitary as
required.

To this end, consider the Choi-Jamio{\l}kowski representation of
$\Psi$, which is
\[
J(\Psi) = \int \op{vec}(U)\op{vec}(U)^{\ast}
\ip{\op{vec}(U)\op{vec}(U)^{\ast}}{J(\Phi)} \mathrm{d}\mu(U).
\]
By the identity \eqref{eq:identity2} we have
\begin{align*}
J(\Psi) & = 
\frac{2}{d(d-1)}
\tr_{\Y_2\otimes\X_2}\left[
\left(R_{\Y_1\otimes\Y_2} \otimes R_{\X_1\otimes\X_2}\right)
\left(\I_{\Y_1\otimes\X_1} \otimes J(\Phi)\right)\right]\\
& \quad
+ \frac{2}{d(d+1)}
\tr_{\Y_2\otimes\X_2}\left[
\left(S_{\Y_1\otimes\Y_2} \otimes S_{\X_1\otimes\X_2}\right)
\left(\I_{\Y_1\otimes\X_1} \otimes J(\Phi)\right)\right],
\end{align*}
where, in this equation, it is viewed that $J(\Phi)\in
\lin{\Y_2\otimes\X_2}$ and $J(\Psi)\in \lin{\Y_1\otimes\X_1}$.
(As in the previous section, the spaces $\X_1$, $\X_2$ and $\Y_1$,
$\Y_2$ are isomorphic copies of $\X$ and $\Y$, respectively.)

The above expression of $J(\Psi)$ may now be simplified by means of
the equations \eqref{eq:R-and-S}.
In particular, we have
\begin{align*}
J(\Psi) & =
\frac{1}{2d(d-1)}
\left[ \tr(J(\Phi)) \I_{\Y\otimes\X} \,-\, \tr_{\X}(J(\Phi))\otimes \I_{\X}
\,-\, \I_{\Y}\otimes \tr_{\Y}(J(\Phi)) \,+\, J(\Phi)\right]\\
& \quad +
\frac{1}{2d(d+1)}
\left[ \tr(J(\Phi)) \I_{\Y\otimes\X} \,+\, \tr_{\X}(J(\Phi))\otimes \I_{\X}
\,+\, \I_{\Y}\otimes \tr_{\Y}(J(\Phi)) \,+\, J(\Phi)\right].
\end{align*}
Making use of the equalities
$\tr_{\X}(J(\Phi)) = \I_{\Y}$ and
$\tr_{\Y}(J(\Phi)) = \I_{\X}$ (which follow from $\Phi$ being doubly
stochastic), we have
\[
J(\Psi) = 
\frac{d^2 - 2}{d(d^2 - 1)} \I_{\Y\otimes\X} + \frac{1}{d^2 - 1}
J(\Phi)
= 
\frac{d^2 - 2}{d^2 - 1} J(\Omega) + \frac{1}{d^2 - 1} J(\Phi).
\]
As $J$ is a linear bijection, this implies that equation
\eqref{eq:Psi} holds, and therefore completes the proof.
\end{proof}

This theorem implies that, within the smallest real affine subspace of
super-operators that contains the doubly stochastic channels, there is
a ball with positive radius around the completely depolarizing channel
within which all super-operators are mixed-unitary.
This fact is established by the following corollary.

\begin{cor} \label{cor:ball}
  Suppose $\Phi:\lin{\X}\rightarrow\lin{\Y}$ is a trace-preserving and 
  unital super-operator such that $J(\Phi)$ is Hermitian and satisfies 
  \[
  \norm{J(\Phi) - \frac{1}{d}\I_{\Y\otimes\X}}_{\infty}
  \leq\frac{1}{d(d^2-1)}.
  \]
  Then $\Phi$ is a mixed-unitary channel.
\end{cor}

\begin{proof}
Define $\Psi = (d^2 - 1)\Phi - (d^2 - 2)\Omega$, which is clearly
trace-preserving and unital.
It follows from the expression
\[
J(\Psi) = (d^2 - 1)\left( J(\Phi) - J(\Omega) \right) + J(\Omega)
=
(d^2 - 1)\left( J(\Phi) - \frac{1}{d}\I_{\Y\otimes\X} \right) +
\frac{1}{d}\I_{\Y\otimes\X},
\]
together with the assumptions of the corollary, that $\Psi$ is
completely positive, and thus is doubly stochastic.
By Theorem~\ref{theorem:main1} we therefore have that
\[
\frac{d^2 - 2}{d^2 - 1}\Omega + \frac{1}{d^2 - 1} \Psi = \Phi
\]
is mixed-unitary, as required.
\end{proof}

\section{Discussion}

As has been discussed by several authors, the 
Choi-Jamio{\l}kowski representation reveals density operator analogues
to various facts about channels.
In the present case, the mapping
\[
\Phi \mapsto \frac{1}{d}J(\Phi)
\]
gives a linear bijection from the set of doubly stochastic
channels $\Phi:\lin{\X}\rightarrow\lin{\Y}$ to the collection
of density operators $\rho\in\density{\Y\otimes\X}$ whose reduced
states on both $\Y$ and $\X$ are completely mixed:
\begin{equation} \label{eq:completely-mixed-reductions}
\tr_{\Y}(\rho) = \frac{1}{d}\I_{\X}
\quad\quad\text{and}\quad\quad
\tr_{\X} (\rho) = \frac{1}{d}\I_{\Y}.
\end{equation}
Corollary \ref{cor:ball} therefore establishes that within the
smallest affine subspace that contains all such density operators,
there is a ball of positive radius around the maximally mixed state
containing only states that are expressible as mixtures of 
{\it maximally entangled} states.
When expressed in these terms, it is appropriate to compare the
results of the previous section with an analogous fact concerning the
set of separable states, which are those expressible as mixtures of 
{\it unentangled} states.

Specifically, it follows from the results proved in \cite{GurvitsB02}
that any unit-trace Hermitian operator $\rho\in\lin{\Y\otimes\X}$
satisfying 
\begin{equation} \label{eq:sep-ball}
\norm{\rho - \I/d^2}_2 \leq \frac{1}{d^2}
\end{equation}
is a separable density operator.
In contrast, Corollary~\ref{cor:ball} establishes that any
unit-trace Hermitian operator $\rho\in\lin{\Y\otimes\X}$
satisfying both the linear constraints
\eqref{eq:completely-mixed-reductions} and the bound 
\begin{equation} \label{eq:max-ball}
\norm{\rho - \I/d^2}_{\infty} \leq \frac{1}{d^2(d^2 - 1)},
\end{equation}
is expressible as a convex combination of maximally entangled states.
(It is clear that operators not satisfying the constraints
\eqref{eq:completely-mixed-reductions} are not expressible as convex
combinations of maximally entangled states, regardless of their
distance from the maximally mixed state.)

It is noted that the ball defined by \eqref{eq:max-ball} is properly
contained in the one defined by \eqref{eq:sep-ball}.
All of the states contained in the smaller ball are therefore
representable at two extremes: as convex combinations of unentangled
states and convex combinations of maximally entangled states.
To what extent this fact can be generalized is an interesting question
for future research.

\subsection*{Acknowledgements}

I thank Andrew Childs, Gus Gutoski, and Marco Piani for comments
and suggestions.
This research was supported by Canada's NSERC and the Canadian
Institute for Advanced Research (CIFAR).


\end{document}